\documentclass[notitlepage,showkeys,nofootinbib,11pt,tightenlines,superscriptaddress
]{revtex4-2}
\usepackage{placeins}
\usepackage{amsmath,comment,subfigure}
\usepackage{placeins}
\usepackage{enumitem}
\usepackage{siunitx,soul}
\usepackage{graphicx}
\usepackage{color,makecell,adjustbox,multirow}
\usepackage{hhline}
\usepackage{setspace}

\usepackage{dcolumn}
\usepackage{bm}
\makeatletter
\let\ps@titlepage\ps@plain
\makeatother
\usepackage[utf8]{inputenc}
\usepackage[T1]{fontenc}
\usepackage{mathptmx,epstopdf} 
\usepackage[colorlinks=true,linkcolor=blue]{hyperref}
\hypersetup{
colorlinks=true,    
linkcolor=blue,       
citecolor=blue,       
filecolor=blue,        
urlcolor=blue,        
linktoc=page          
}

\begin{document}
\onecolumngrid
\title{Designing Pr-based Advanced Photoluminescent Materials using Machine Learning and
Density Functional Theory}

\author{Upendra Kumar $^\#$}
\affiliation{Center of Material Digitalization, Korea Institute of Ceramic Engineering and Technology (KICET), Jinju 52851, South Korea.}

\author{Hyeon Woo Kim $^\#$}
\affiliation{Center of Material Digitalization, Korea Institute of Ceramic Engineering and Technology (KICET), Jinju 52851, South Korea.}
\affiliation{Department of Materials Science and Engineering, Hanyang University, Seoul 04763, South Korea.}
\author{Sobhit Singh}
\affiliation{ Department of Mechanical Engineering at the University of Rochester, New York 14611, United States.}
\author{Hyunseok Ko}
\email{hko@kicet.re.kr}
\affiliation{Center of Material Digitalization, Korea Institute of Ceramic Engineering and Technology (KICET), Jinju 52851, South Korea.}
\author{Sung Beom Cho}
\email{csb@ajou.ac.kr}
\affiliation{Department of Energy Systems Research, Ajou University, Suwon 16499, South Korea.}


\def\thefootnote{$^\#$}\footnotetext{These authors contributed equally to this work}\def\thefootnote{\arabic{footnote}}

\begin{abstract}
\noindent This work presents a machine learning  approach to predict novel perovskite oxide materials in the Pr-Al-O and Pr-Sc-O compound families with the potential for photoluminescence applications. The predicted materials exhibit a large bandgap and high Debye temperature, and have remained unexplored thus far. The predicted compounds (Pr$_3$AlO$_6$, Pr$_4$Al$_2$O$_9$, Pr$_3$ScO$_6$ and Pr$_3$Sc$_5$O$_{12}$) are screened using machine learning approach, which are then confirmed by density functional theory calculations. The study includes the calculation of the bandgap and density of states to determine electronic properties, and the optical absorption and emission spectra to determine optical properties. Mechanical stability of the predicted compounds, as demonstrated by satisfying the Born-Huang criterion. By combining machine learning and density functional theory, this work offers a more efficient and comprehensive approach to materials discovery and design.
\end{abstract}
\maketitle
\vspace{-0.8cm}

\section{Introduction}
Luminescent materials based on perovskite halides, such as methylammonium lead iodide (MAPbI$_3$), have shown great promise for use in photovoltaic and optoelectronic applications \cite{kim2014role,hutter2017direct}. However, the stability of these materials remains a significant challenge, particularly due to their sensitivity to moisture and oxygen at ambient conditions \cite{li2020towards}. To address this issue, a range of strategies have been explored, including the use of encapsulation \cite{liu2016enhanced} and protective coatings \cite{ye2023stabilization}. However, such approaches can be complex and costly, thereby limiting the practical applicability of perovskite halides. In addition to exploring encapsulation and protective coatings, researchers have also investigated a variety of other materials in an effort to develop more stable luminescent materials. Despite these efforts, the stability of perovskite halides remains a key challenge in their practical implementation.
\par Furthermore, a majority of perovskite oxides exhibit rigid stability on humidity, however, they are not considered as common photovoltaic materials due to their wide band-gap energies. Nonetheless, some of perovskite oxides shows excellent showcasing distinct optical properties. These properties are utilized in the development of advanced optoelectronic devices, such as nonlinear optics crystals \cite{kaminow1967quantitative,zhu1997quasi}, scintillators \cite{wojtowicz2006scintillation}, photoluminescent (PL) and electroluminescent materials \cite{kan2005blue,takashima2009low}, as well as solar cells \cite{yang2010above}. Novel PL bands have been observed in SrTiO$_3$, which is a typical example of a perovskite semiconductor \cite{yasuda2008dynamics,yamada2009temperature}. Perovskite oxide derivatives offer an attractive alternative to perovskite halides for luminescent applications \cite{song2019luminescent}. These materials are often more thermally stable \cite{leijtens2017towards} and less toxic \cite{jiang2021tin} than their halide counterparts, and can exhibit desirable electronic and optical properties. Furthermore, perovskite oxide derivatives provide a diverse design space that allows for the incorporation of traditional luminescent elements such as Cr, Yb, Pr, Eu, Tb, and others. The chemical space of perovskite oxide derivatives is not fully explored yet, offering a promising avenue for the development of new luminescent materials.
\par Perovskite oxides are promising scintillators due to their high light yield and fast response time \cite{liu2020band}. They emit more photons per unit of absorbed radiation than other materials, making them useful for detecting high-energy particles and reducing the risk of radiation damage to sensitive equipment \cite{kucera2022scintillation}. Additionally, perovskite oxide scintillators have the potential to overcome the stability issues of perovskite halide scintillators. While perovskite halide scintillators have high light yield and fast response times, they are known to be unstable under certain conditions such as exposure to moisture and high temperatures. Several types of perovskite oxide scintillators have been studied, including strontium titanate (SrTiO$_3$) \cite{sohrabi2015photoluminescence}, barium titanate (BaTiO$_3$) \cite{singh2017intense}, and lanthanum aluminate (LaAlO$_3$) \cite{pejchal2022untangling}, which have shown promising results in terms of their light yield and response time. Therefore, ongoing research is focused on improving their performance and understanding their underlying physics.
\par In this study, we have identified under-explored Pr-based perovskite oxides and predicted new promising compounds for photovoltaic applications. By combining machine learning and data mining, we found that Pr-based perovskites are relatively under-explored compared to other types. The focus of this work is on predicting perovskite oxide materials in the Pr-Al-O and Pr-Sc-O families, known for their larger bandgap and Debye temperature, as potential candidates for photoluminescence applications. Machine learning is employed to screen a vast number of materials and predict their electronic and optical properties. To validate the predictions, density functional theory (DFT) calculations are performed to study the band structure, density of states, optical absorption, emission spectra, as well as elastic and mechanical stability. This integrated approach of machine learning and DFT offers a more efficient and comprehensive method for materials discovery and design, facilitating the identification of perovskite oxide materials with desirable electronic and optical properties for photoluminescence applications.
\section{Methods}
To construct the ML models, we employed convolutional neural networks (CGCNN) \cite{CGCNN} method. The core concept behind the CGCNN method \cite{CGCNN} is to first represent a crystal structure using a crystal graph that encodes both atomic information and bonding interactions between atoms. Then design a convolutional neural network on top of the graph in order to automatically extract representations that are optimal for predicting target properties by training  the ML model with DFT calculated data.  
\par The Vienna ab initio simulation package (VASP), which uses the projector augmented wave method (PAW), is used to perform all the reported DFT calculations in this work \cite{kresse1996efficiency,kresse1996efficient}. Exchange and correlation energies are computed using the generalized gradient approximation provided as parameterized by Perdew-Burke-Ernzerhof (PBE) \cite{PBE}.
Brillouin zone is constructed using a $\Gamma$-centered (3$\times$3 $\times$5 for Pr$_3$AlO$_6$ and 2$\times$3 $\times$2 for Pr$_4$Al$_2$O$_9$) and Monkhorst-Pack (4$\times$4$\times$4 for Pr$_3$ScO$_6$ and  2$\times$2 $\times$2 for Pr$_3$Sc$_5$O$_{12}$) type k-meshes \cite{Monkhorst}. In all DFT calculations, we employed a plane wave cut-off energy of 400 eV. We used \mbox{$10^{-4}$ eV$/\textup{~\AA}$}  as the force convergence criterion to relax the inner-atomic coordinates, and
10$^{-6}$ eV as the energy convergence criterion for self-consistence DFT calculations.  

\section{Result and Discussion}
\par  The Debye temperature ($\Theta_{\mathrm{D}}$) is the maximum temperature that can be attained as a result of a single normal-mode vibration i.e. the temperature of a crystal's highest normal mode of vibration. It is a good indicator of structural stiffness, which makes it suitable for evaluating photoluminescent quantum yield \cite{denault2014consequences}. However, there are a number of limitations to utilizing the DFT method to compute  $\Theta_{\mathrm{D}}$ i.e. doing so is computationally expensive. Instead, it is also possible to predict $\Theta_{\mathrm{D}}$ for many compounds using machine learning \cite{zhuo2018identifying}, which is computationally less expensive than DFT. Still, knowing only the $\Theta_{\mathrm{D}}$ of a crystal structure is insufficient to achieve a high photoluminescence quantum efficiency: \textit{a wide bandgap is also required}. Plotting $\Theta_{\mathrm{D}}$ as a function of DFT determined bandgap, which serves as a sorting diagram, allows for the final optimization of these two features. There is a dependency of  $\Theta_{\mathrm{D}}$ on the energy band gap $(E_{g})$ in the semiconducting material \cite{ullrich2017relation}. Therefore, it is necessary to find the value of bandgap and $\Theta_{\mathrm{D}}$ for a good photoluminescence material. \par
\begin{figure}[h]
\centering
\hspace{-0.4cm}\includegraphics[width=0.90\linewidth]{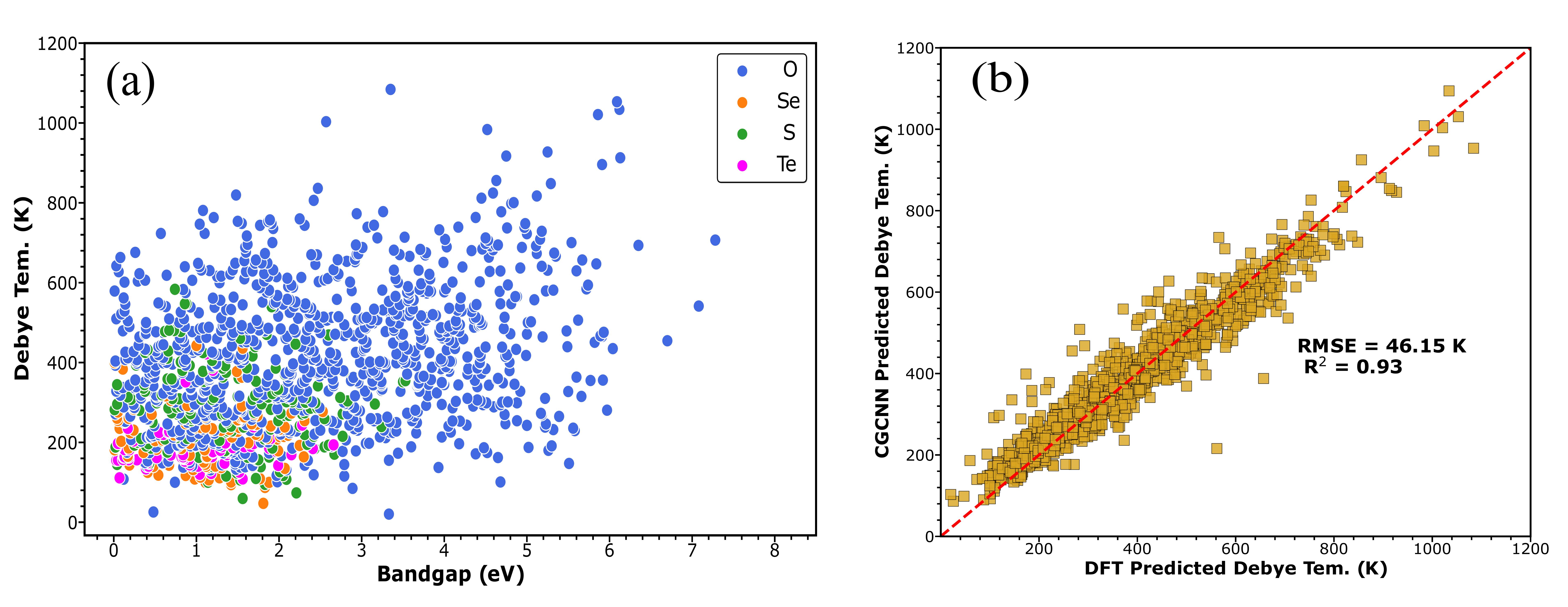}
\caption{\small \textbf{(a)} Plot between Debye temperature and bandgap of chalcogenide perovskites family. \textbf{(b)} Plot between DFT (PBE) predicted vs CGCNN predicted values of Debye temperature.}
\label{Debye_BandGap}
\end{figure}

 Advanced ML-based methods have revolutionized the field of material science by providing an alternative to traditional experimental trial-and-error and computationally expensive DFT calculation techniques \cite{agrawal2016perspective}. In the Pearson’s crystal database (PCD)\cite{villars2007pearson}, the machine learning has been used to predict $\Theta_{\mathrm{D}}$ for a majority of compounds. Therefore, we performed data mining for O, Se, S and Te based chalcogen ternary compounds by using  materials project \cite{jain2013commentary} database, as shown in Fig.\ref{Debye_BandGap}\textbf{(a)}. It is found that oxide-based materials consist  large bandgap and high Debye temperature among the studied chalcogen family. Since data availability for Debye temperature is very limited in the materials-project database \cite{jain2013commentary},
we employed CGCNN \cite{CGCNN} for the calculation of calculating Debye temperature with the crystallographic
information files as an input feature.
The CGCNN is able to predict the Debye temperature with accuracy of 93\% as shown in Fig.\ref{Debye_BandGap}\textbf{(b)}.
\par
\begin{figure}[h]
\centering
\includegraphics[width=.90\linewidth]{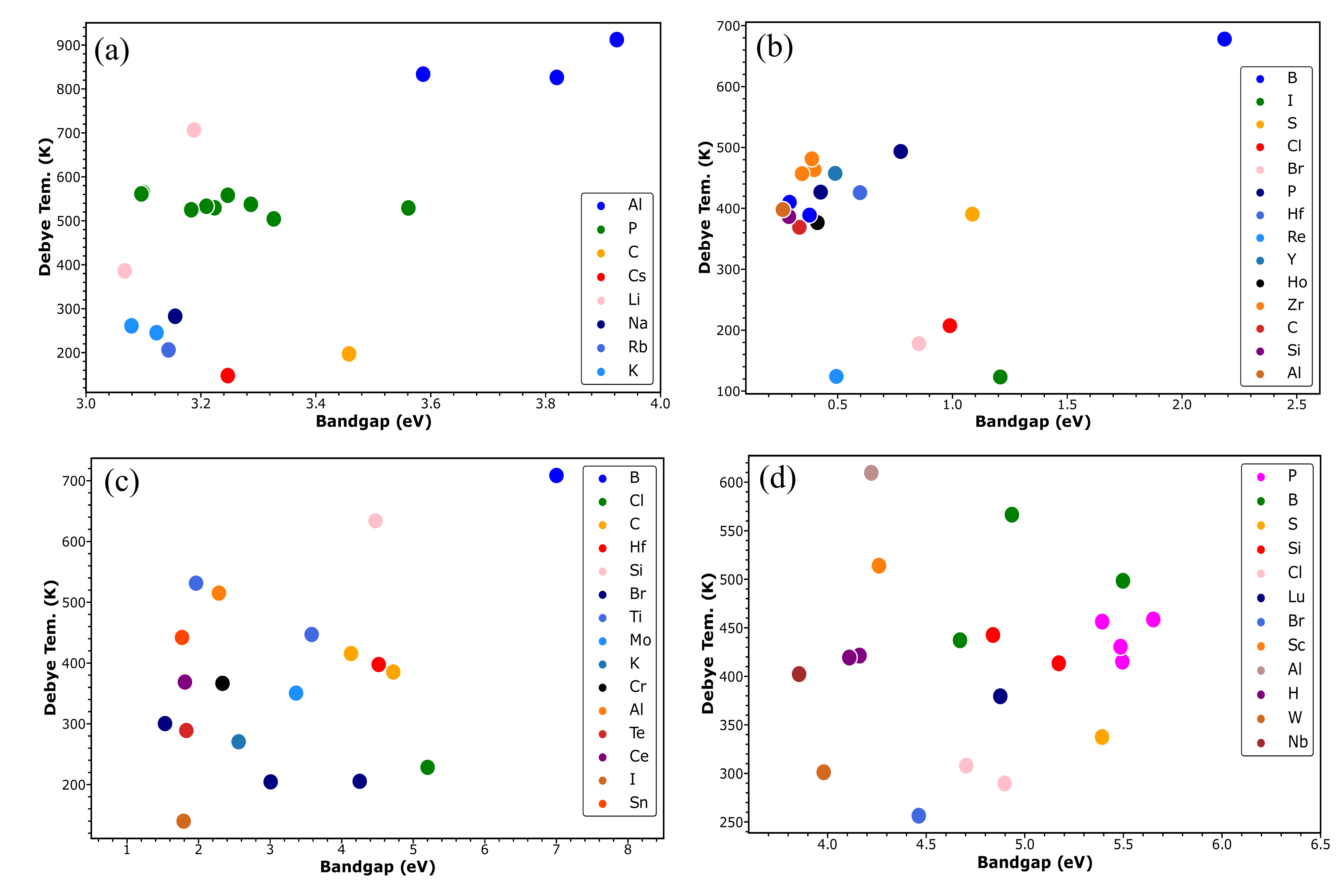}
\caption{\small Plot between Debye temperature and bandgap of \textbf{(a)} Cr-based \textbf{(b)} Eu-based \textbf{(c)} Yb-based, and \textbf{(d)} Pr-based ternary perovskites oxide family.}
\label{Cr_Eu_Yb_PrO_Family}
\end{figure}

   We started our data mining process using the materials-project database \cite{jain2013commentary} to explore the relatively under-explored perovskite oxide family containing Cr, Eu, Yb, and Pr elements, as shown in Fig.(\ref{Cr_Eu_Yb_PrO_Family}), with further details provided in {\color{blue}{supplementary section (I)}}. After screening several promising photoluminescent materials using data mining, we found that Praseodymium (Pr) is the least explored element in this family. Therefore, we focused on the ternary compounds of the Pr based perovskite oxide family, which have a large bandgap and high Debye temperature, as depicted in Fig.\ref{Cr_Eu_Yb_PrO_Family}\textbf{(d)}. Among the Pr based perovskite oxide family, the \mbox{Pr-Sc-O} and \mbox{Pr-Al-O} subfamilies are the least explored. We found only two well-known materials having zero energy above the  convex hull (E$_\mathrm{hull}$) in the materials-project database \cite{jain2013commentary}, as described in {\color{blue}{supplementary section (I:D)}} \cite{jain2013commentary}. Hence, we explored these families to predict new candidate structures for potential photoluminescence applications.

\par
\begin{figure}
\centering
\includegraphics[width=.65\linewidth]{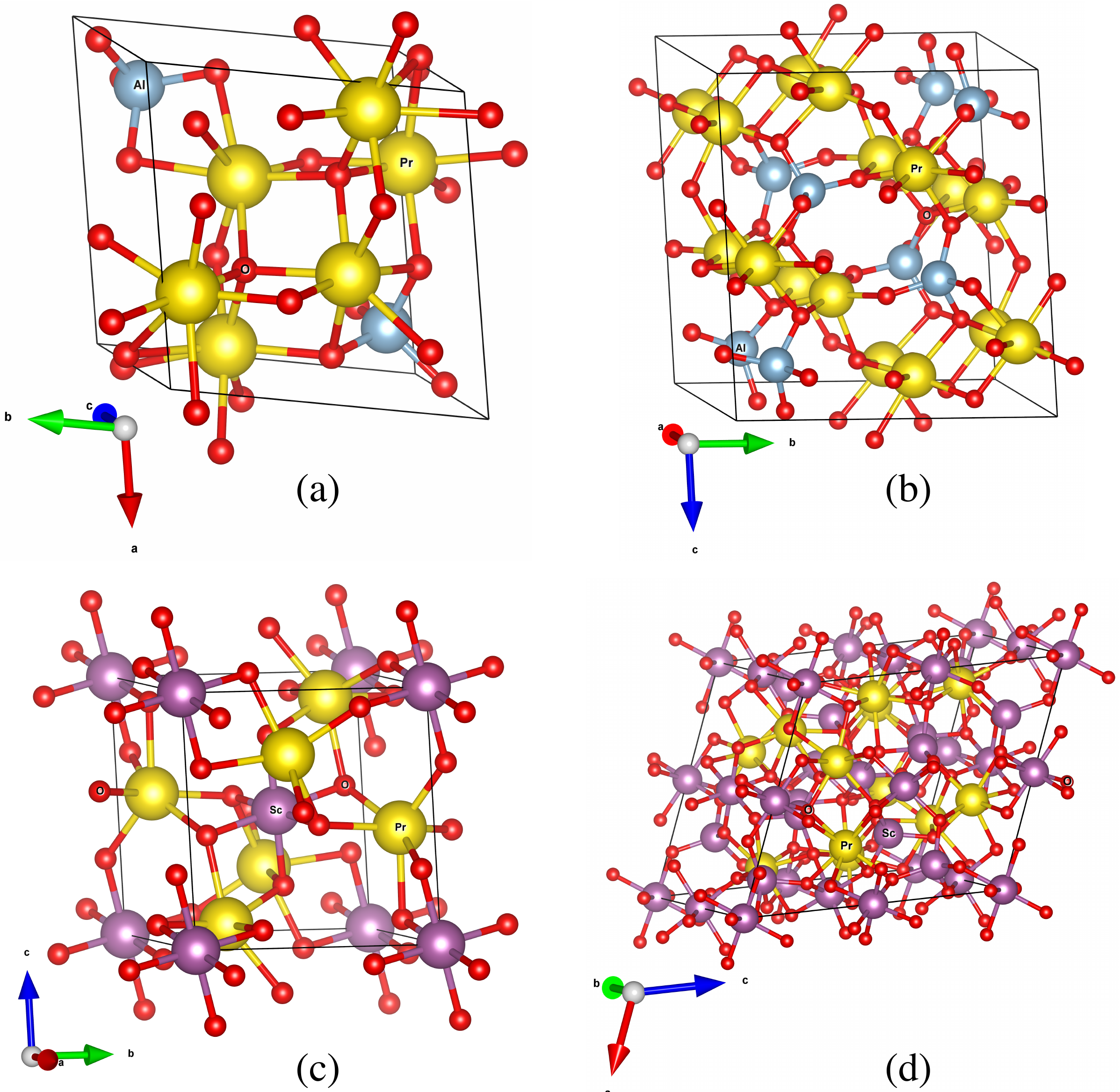}
\caption{\small Crystal structures of \textbf{(a)}  Pr$_3$AlO$_6$ ($Cmc2_1$), \textbf{(b)} Pr$_4$Al$_2$O$_9$ ($P2_1/c$), \textbf{(c)} Pr$_3$ScO$_6$ ($R\bar{3}$), and \textbf{(d)} Pr$_3$Sc$_5$O$_{12}$ ($Ia\bar{3}d$)..}
\label{ball_stick}
\end{figure}

 We have used the substitution method to search for more crystal structures. A machine learning model for ionic substitution based on experimental data has already been proposed by Hautier et al. \cite{hautier2011data}. In this work, we utilize the same machine learning model for the prediction of new structures in Pr-Al-O and Pr-Sc-O ternary compound families. In the case of Pr-Al-O family, there are only three obtained compositions,
i.e., \textbf{(i)} Pr$_4$Al$_2$O$_9$ \textbf{(ii)} PrAlO$_3$ and \textbf{(iii)} Pr$_3$AlO$_6$  having zero energy above the convex hull. We note, PrAlO$_3$  has already been in the category of well known compound \cite{jain2013commentary}. Therefore, there are only two remaining candidates  i.e.  \textbf{(i)} Pr$_4$Al$_2$O$_9$ and \textbf{(ii)} Pr$_3$AlO$_6$, which can be considered as novel candidates in this work. Similarly, in the case of Pr-Sc-O family, the compounds Pr$_3$ScO$_6$ and Pr$_3$Sc$_5$O$_{12}$ are considered as novel candidates in this work. All the details of newly predicted compounds are given in Table(\ref{New_Compound}). Their structures are shown in the Fig.(\ref{ball_stick}). 
\begin{table}[h]
\centering
\scalebox{0.62}{\begin{tabular}{|c|c|c|c|c|c|c|}
\hline
& \thead{\textbf{Lattice }\\ {\textbf{Parameters (\AA)}}}&\thead{\textbf{Lattice}\\ {\textbf{Angle }}} & \thead{\textbf{Space Group}\\ {\textbf{Type}}}  & \thead{\textbf{Crystal }\\ {\textbf{Structure}}}&\thead{\quad \textbf{Space Group} \quad\\ {\quad \textbf{Number}}}\quad \\
\hline
&&&&&\\
\thead{{}\\Pr$_3$AlO$_6$ \\ {}} \quad & \quad \thead{a = 7.47  \\
b = 7.47 \\
c = 5.62} \quad & \quad \thead{$\alpha$ = 90$^\circ$ \\
$\beta$ = 90$^\circ$ \\
$\gamma$ = 102.80$^\circ$ } \quad&  \quad \thead{{}\\$Cmc2_1$ \\ {}}&\quad \thead{{}\\Orthorhombic \\ {}}& \thead{36 \\ {}} \\
&&&&&\\
\hline
&&&&&\\
\thead{{}\\Pr$_4$Al$_2$O$_9$ \\ {}}& \thead{a = 11.02\\
b = 7.85 \\
c = 11.50} & \thead{$\alpha$ = 70.54$^\circ$\\
$\beta$ = 90$^\circ$\\
$\gamma$ = 90$^\circ$} & \thead{{}\\$P2_1/c$\\ {}} & \thead{{}\\Monoclinic\\{}}&\thead{14 \\ {}}\\
&&&&&\\
\hline
&&&&&\\
\thead{{}\\Pr$_3$ScO$_6$ \\ {}}& \thead{a = 6.89 \\
b = 6.89 \\
c = 6.89}  & \thead{$\alpha$ = 92.38$^\circ$ \\
$\beta$ = 92.38$^\circ$ \\
$\gamma$ = 92.38$^\circ$ } & \thead{{}\\$R\bar{3}$ \\ {}}& \thead{ {Trigonal}\\ (Rhombohedral) \\ {}}&\thead{148 \\ {}} \\
&&&&&\\
\hline
&&&&&\\
\thead{{}\\Pr$_3$Sc$_5$O$_{12}$ \\ {}}& \thead{a =11.39 \\
b = 11.39 \\
c = 11.39} & \thead{$\alpha$ = 109.47$^\circ$ \\
$\beta$ = 109.47$^\circ$\\
$\gamma$ = 109.47$^\circ$} & \thead{{}\\$Ia\bar{3}d$ \\ {} }& \thead{{}\\Cubic \\ {}}& \thead{230\\ {}}\\
&&&&&\\
\hline
\end{tabular}}
\caption{\small The structural details of newly predicted compounds.}
\label{New_Compound}
\end{table}
\par
\begin{figure}[h]
\centering
\includegraphics[width=0.9\linewidth]{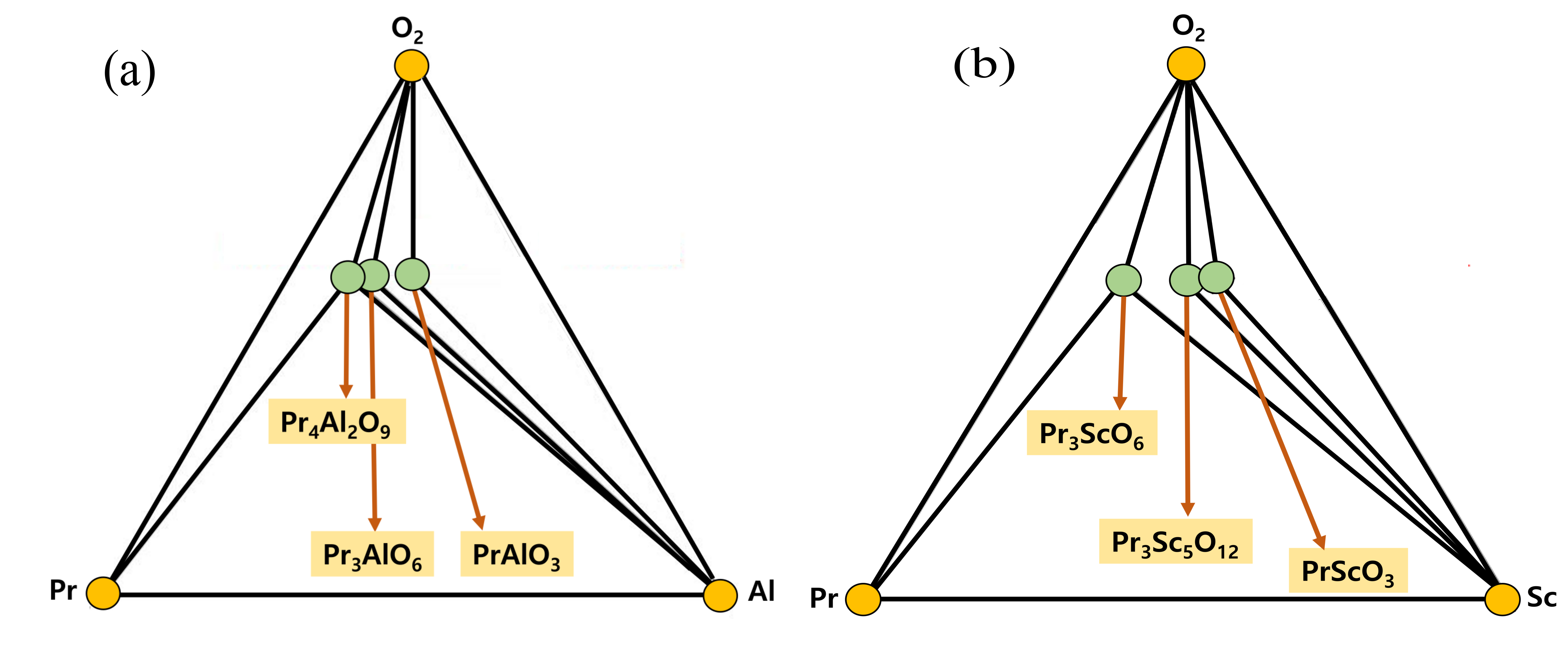}
\caption{\small The convex hull plot of \textbf{(a)} Pr-Al-O \textbf{(b)} Pr-Sc-O family. The PrAlO$_3$ and PrScO$_3$ represent known compounds of the materials-project database \cite{jain2013commentary}, and others are  newly predicted  compounds by our ML combined with DFT method.}
\label{convex_hull_plot}
\end{figure}

 \noindent Getting the formation energy and building the related convex hull is a crucial step in figuring out if a compound is energetically stable  \cite{peterson2021materials}. From thermodynamical point of view, the convex hull belongs to the Gibbs free energy of the compounds at zero temperature. Our calculations reveal that the newly predicted compounds are energetically stable, as confirmed by the convex hull plot shown in Fig.(\ref{convex_hull_plot}). Further details, including E$_{\mathrm{hull}}$ and parent atom details, are provided in {\color{blue}{supplementary section (II)}}. All these predicted compounds are mechanically stable as discussed in more detail in the {\color{blue}{supplementary section(III). }}
\\
\subsection{Bandgap} We calculated electronic bandgap of the newly predicted compounds using PBE-DFT~\cite{PBE}, CGCNN, and modified Becke Johnson (mBJ)~\cite{becke2006simple} methods.
For further improving the accuracy of the bandgap, we utilized computationally more expensive hybrid functional -- Heyd-Scuseria-Ernzerhof (HSE06) \cite{heyd2005energy}. 
All the calculated bandgap values are reported in  the Table(\ref{BandGapDebye}).
We have also calculated Debye temperature of the newly predicted compounds using CGCNN \cite{CGCNN} and validated it with DFT calculations,  as mentioned in the Table(\ref{BandGapDebye}). All the predicted compounds have a large bandgap and high Debye temperature.
\begin{table}[h]
\centering
\scalebox{0.95}{\begin{tabular}{|c|c|c|c|c|c|c|c|}
\hline
& \thead{\textbf{Bandgap}\\ {\textbf{DFT-PBE (eV)}}}&\thead{\textbf{Bandgap}\\ {\textbf{CGCNN (eV)}}} & \thead{\textbf{Bandgap}\\ {\textbf{DFT-mBJ (eV)}}}  &  \thead{\textbf{Bandgap}\\{\textbf{DFT-HSE (eV)}}} & \thead{\textbf{Debye Temp.}\\\textbf{CGCNN (K)}} & \thead{\textbf{Debye Temp.}\\ \textbf{DFT-PBE (K)}} & \thead{\textbf{Bandgap} \\ \textbf{Type}}\\
\hline
&&&&&&&\\
Pr$_3$AlO$_6$ \quad & \quad 4.18 \quad & \quad 4.18\quad&  \quad 5.69 \quad &  \quad 5.67 \quad &\quad  393.98 \quad &  \quad 398.78 \quad& Indirect\\
&&&&&&& \\
\hline
&&&&&&&\\
Pr$_4$Al$_2$O$_9$& 4.07 & 3.84 & 6.07 & 5.56  & 428.87 & 431.33& Direct  \\
&&&&&&&\\
\hline
&&&&&&&\\
Pr$_3$ScO$_6$& 4.29  & 4.15& 5.46 &  5.80 & 410.60 & 390.24 & Indirect \\
&&&&&&&\\
\hline
&&&&&&&\\
Pr$_3$Sc$_5$O$_{12}$& 3.83& 3.44& 5.12 &  5.32  & 488.55 & 486.34& Direct \\
&&&&&&&\\
\hline
\end{tabular}}
\caption{\small Bandgap and Debye temperature of newly predicted compounds. The DFT-PBE data is used for training the CGCNN model.}
\label{BandGapDebye}
\end{table}
\begin{figure}[h]
\centering
\includegraphics[width=0.6\linewidth]{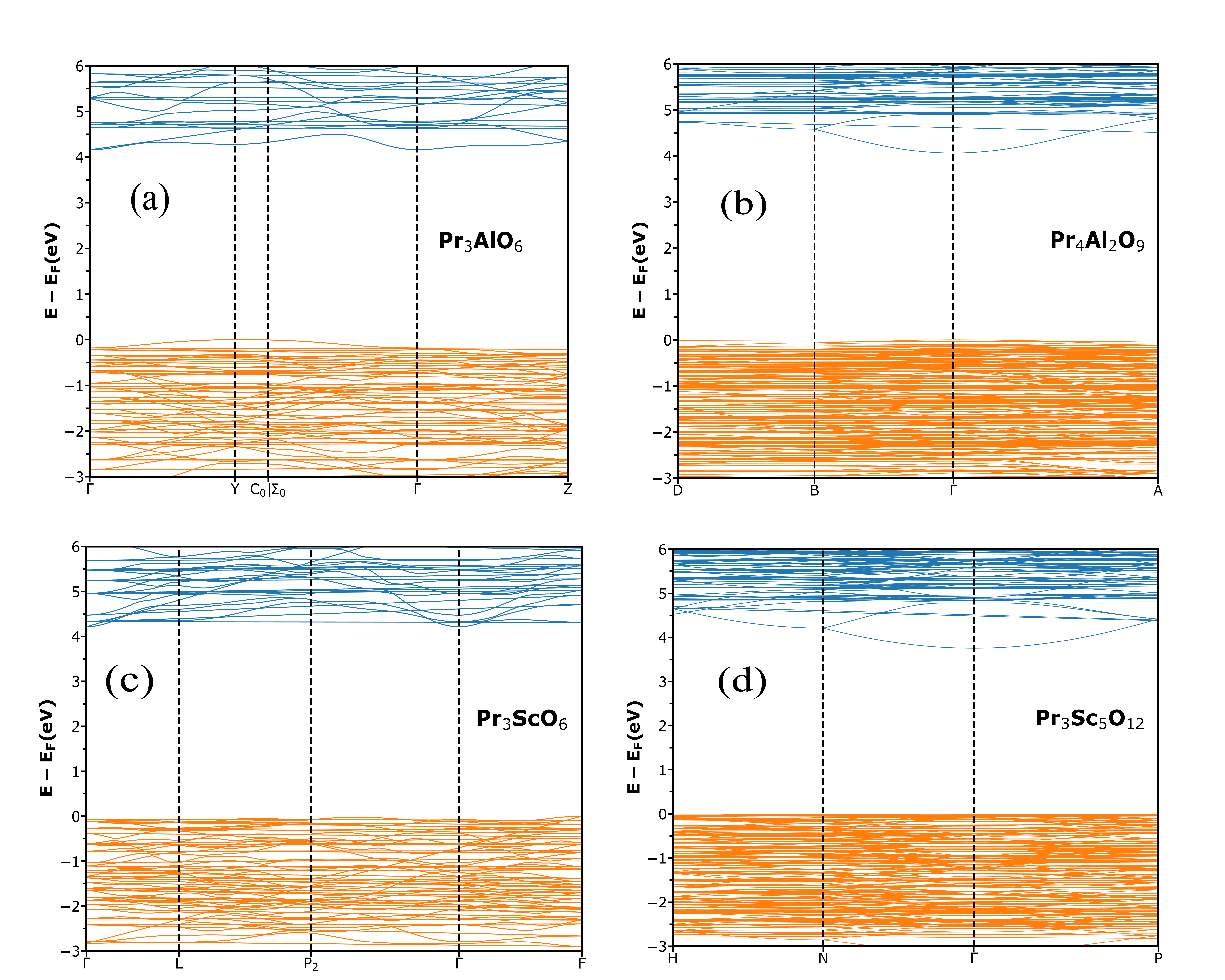}
\caption{\small PBE band structure of \textbf{(a)} Pr$_3$AlO$_6$ \textbf{(b)} Pr$_4$Al$_2$O$_9$ \textbf{(c)} Pr$_3$ScO$_6$ and \textbf{(d)} Pr$_3$Sc$_5$O$_{12}$.
}
\label{bandcollage}
\end{figure}
\par \noindent Direct band gap semiconductors allow for efficient production of photons without assistance from phonons due to the aligned valence and conduction band extrema. This makes them highly desirable for optical devices i.e. for photoluminescence applications. So, the newly predicted compounds Pr$_4$Al$_2$O$_9$ and Pr$_3$Sc$_5$O$_{12}$ are better suited for photoluminescence applications, as shown in Fig.(\ref{bandcollage}). 
\subsection{Density of States (DOS)} 
\begin{figure}[h]
\centering
\includegraphics[width=0.80\linewidth]{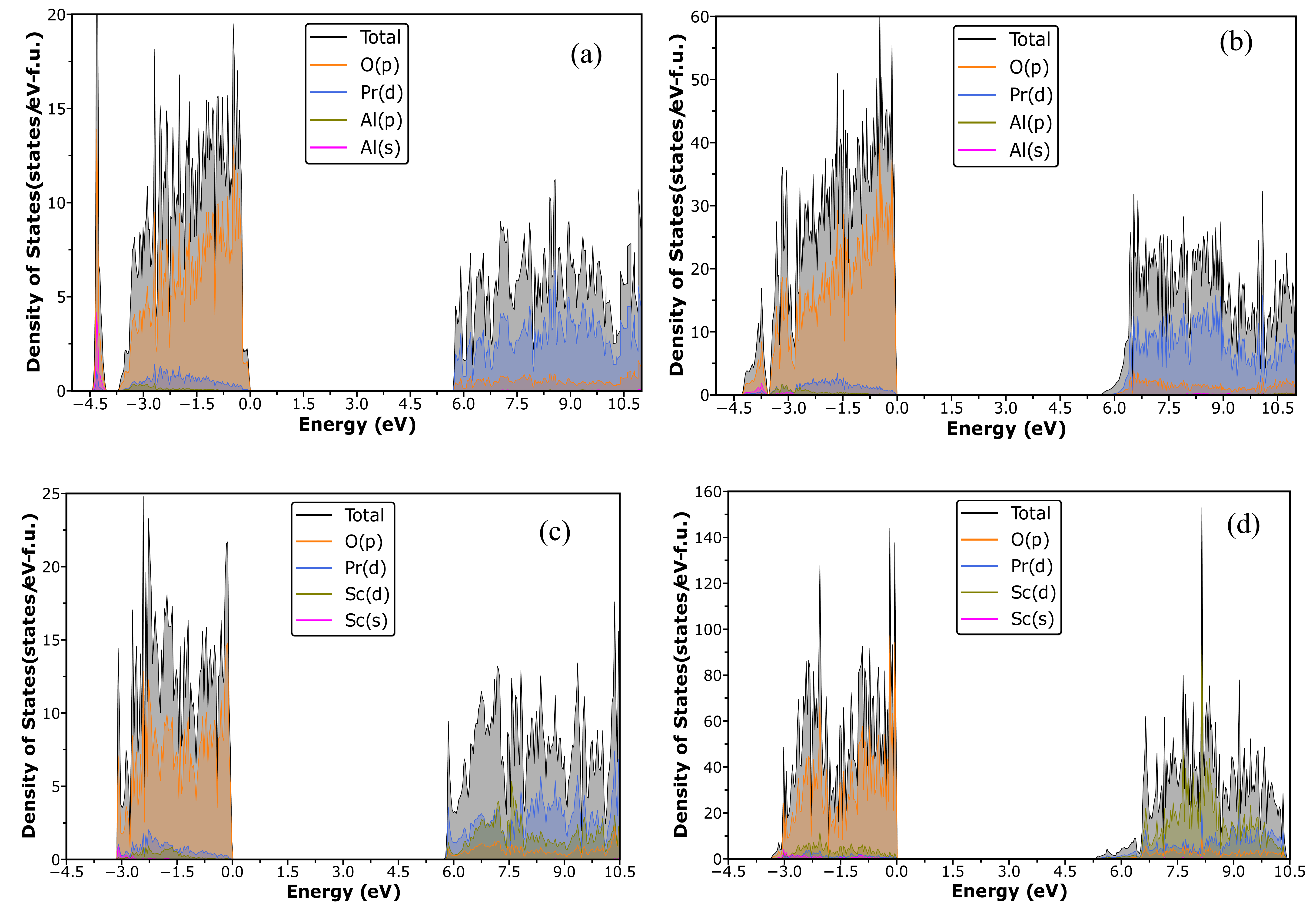}
\caption{\small Total HSE06 electronic density of states and orbital projected density of states for \textbf{(a)} Pr$_3$AlO$_6$ \textbf{(b)} Pr$_4$Al$_2$O$_9$ \textbf{(c)} Pr$_3$ScO$_6$, and \textbf{(d)} Pr$_3$Sc$_5$O$_{12}$.}
\label{dos}
\end{figure}

DOS plays a crucial role in defining the characteristics of the materials \cite{martin2020electronic}. In order to learn more about the electronic structure of the predicted compound, we have calculated the total and atomic-orbitals resolved electronic DOS, as shown in the Fig.(\ref{dos}), respectively. In predicted compounds major contribution is because $p$ orbital of oxygen, $d$ orbital of Pr and Sc.  In DOS of Pr$_3$AlO$_6$, shown in the Fig.\ref{dos}\textbf{(a)}, the peak around -4.3 eV is due to the hybridization of all elements. In case of valance band maximum major contribution in the total DOS is due to $p$ orbital of oxygen. But in case of conduction band minimum, the $d$ orbital of Pr has a major contribution to the total DOS. The $s$ orbital of Al is also playing a role in making larger peak around -4.3 eV. Similar behaviour can be also seen for the  Pr$_4$Al$_2$O$_9$, depicted in the Fig.\ref{dos}\textbf{(b)}. But, here, $s$ orbital of Al contribution can be seen around -4 eV. So, electronic transitions from O$-p$ orbitals to Pr$-d$ orbitals are possible.
Similar pattern of DOS can be seen for the Pr$_3$ScO$_6$ and Pr$_3$Sc$_5$O$_{12}$ compound, depicted in Fig.\ref{dos}\textbf{(c)} and Fig.\ref{dos}\textbf{(d)}.
\subsection{Optical Spectra}
It is possible to derive the linear optical properties from the frequency-dependent complex dielectric function $\epsilon(\omega)$:
\begin{align}
\epsilon(\omega) =\epsilon_{1}(\omega)+i \epsilon_{2}(\omega).
\end{align}
where the real and imaginary components of the dielectric function are denoted by $\epsilon_{1}(\omega)$ and $\epsilon_{2}(\omega)$, respectively, $\omega$ represents the frequency of the photon. The real components $\epsilon_{1}(\omega)$ can be obtained by using the Kramers–Kr\"{o}nig relationship \cite{toll1956causality} and imaginary components $\epsilon_{2}(\omega)$ can be calculated by using momentum matrix elements between the valence and conduction wave functions \cite{ehrenreich1959self}. With the help of $\epsilon_{1}(\omega)$ and $\epsilon_{2}(\omega)$, the refractive index $n(\omega)$ and absorption coefficient $\alpha(\omega)$ can be calculated by using formula:
\begin{align}
n(\omega) = \left[\frac{\sqrt{\epsilon^2_{1}+\epsilon^2_{2}}+\epsilon_{1}}{2}\right]^{\frac{1}{2}},\; \mathrm{and}
\end{align}
\begin{align}
\alpha(\omega) =\sqrt{2}\omega\left[\frac{\sqrt{\epsilon^2_{1}+\epsilon^2_{2}}-\epsilon_{1}}{2}\right]^{\frac{1}{2}}.
\end{align}
\begin{figure}[h]
\centering
\includegraphics[width=0.90\linewidth]{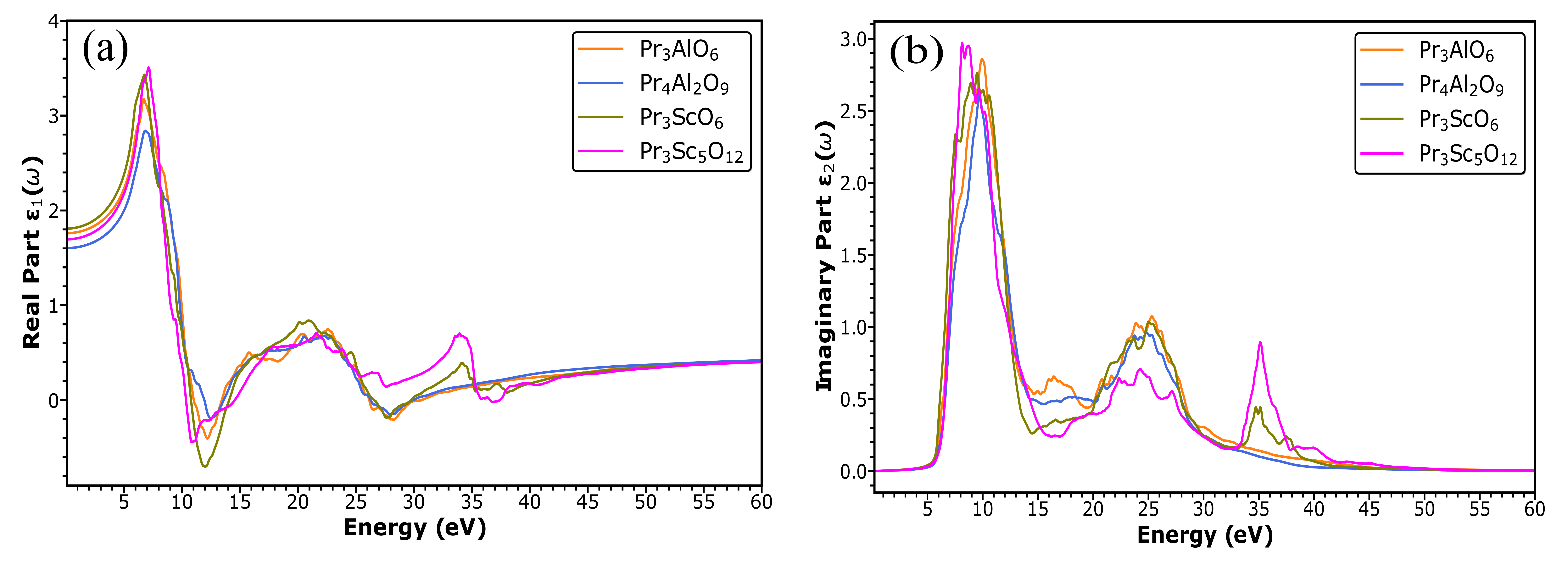}
\caption{\small HSE06 calculated \textbf{(a)} real part  $\epsilon_{1}(\omega)$ and \textbf{(b)} imaginary part $\epsilon_{2}(\omega)$ of the complex dielectric function. }
\label{Real_Imag}
\end{figure}
\par \noindent  
For predicted anticipated compounds, the computed $\epsilon_{1}(\omega)$ and  $\epsilon_{2}(\omega)$ as a function of $\omega$ are shown in Fig.(\ref{Real_Imag}). The maximum peak of real part can be seen at 6.73 eV ($\epsilon_{1}(\omega)=3.17$) for Pr$_3$AlO$_6$, 6.84 ($\epsilon_{1}(\omega)=2.84$) eV for Pr$_4$Al$_2$O$_9$, 6.81 eV ($\epsilon_{1}(\omega)=3.43$) for Pr$_3$ScO$_6$ and 7.13 eV ($\epsilon_{1}(\omega)=3.51$) for Pr$_3$Sc$_5$O$_{12}$ compounds. The zero frequency limits ($\omega \rightarrow 0$) of $\epsilon_{1}(\omega)$ could be used to calculate the static dielectric constants. Static dielectric constants are determined to be 1.75 for Pr$_3$AlO$_6$, 1.60 for Pr$_4$Al$_2$O$_9$, 3.43 for Pr$_3$ScO$_6$, and 1.69 for Pr$_3$Sc$_5$O$_{12}$. The $\epsilon_{2}(\omega)$ in Fig.\ref{Real_Imag}\textbf{(b)} shows that the dielectric function's threshold energy is at about 5.1 eV. This is equivalent to the fundamental absorption edge, which is the optical transition between the valence band maximum (VBM) and the conduction band minimum (CBM). The absorptive portion of $\epsilon_{2}(\omega)$ displays two dominating peaks at nearly 10 eV and 25 eV as the energy increases. The first peak is caused by O$-2p$ electrons transitioning into the $s$ states of cations, but the subsequent peak may correspond to O$-2p$ electrons transitioning into the $p$ states of cations \cite{wang2014structural}.
\begin{figure}[h]
\centering
\includegraphics[width=0.90\linewidth]{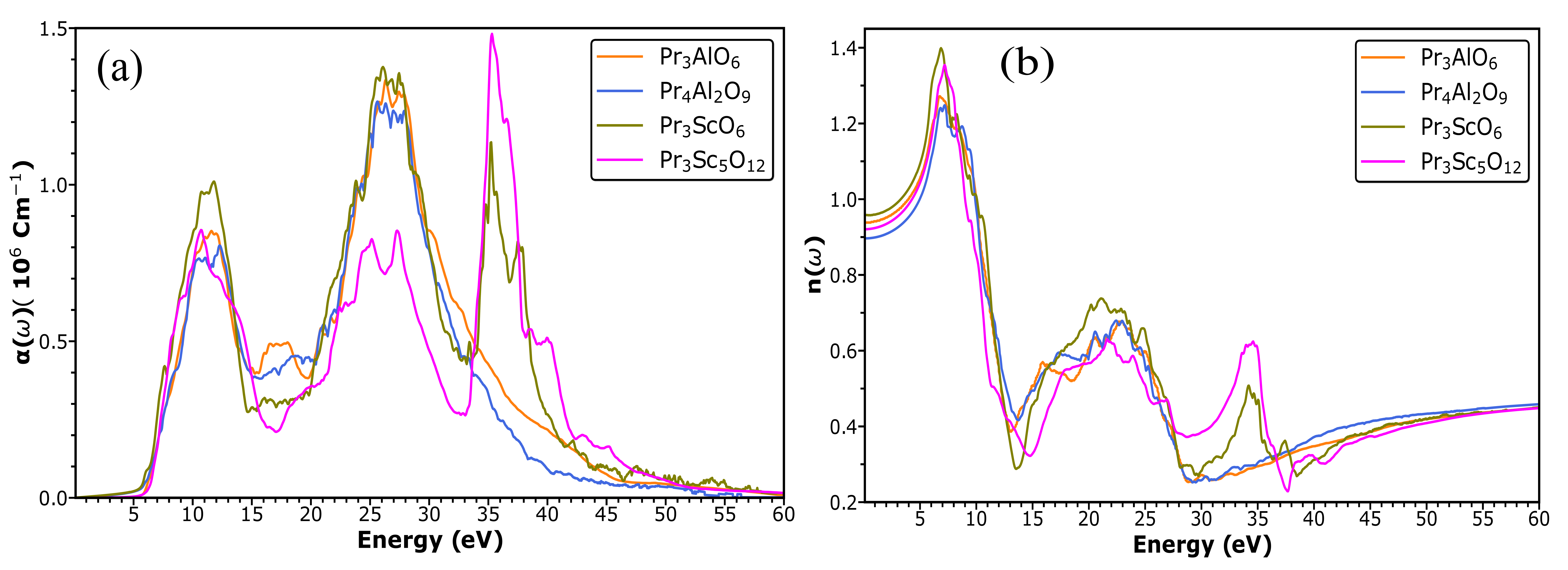}
\caption{\small HSE06 calculated \textbf{(a)} absorption  $\alpha(\omega)$ and \textbf{(b)} refractive $n(\omega)$ spectrum. }
\label{Abs_Refract}
\end{figure}
\par The decay of light intensity propagating over a unit distance in a material is described by the absorption coefficient $\alpha(\omega)$.
According to Fig.\ref{Abs_Refract}\textbf{(a)}, the absorption edge begins to appear at around 5 eV. This is caused by excited electrons transitioning from O$-2p$ states at the top of the valence band to empty cation $2s$ states. Take note that, the $\alpha(\omega)$ is seen at a value lower than $5 $ eV i.e. ultraviolet range. On the other hand, these compounds display a noticeable absorption due to the fact that the absorption coefficient rapidly increases when the photon energy is greater than the absorption edge. This is a property that is typical of semiconductors and insulators.
In Fig.\ref{Abs_Refract}\textbf{(b)}, the measured curve of $n(\omega)$ as a function of photon energy is shown. It should be noted that the static refractive index $n(0)$ values for incoming light are 0.94 for Pr$_3$AlO$_6$, 0.89 for Pr$_4$Al$_2$O$_9$, 0.95 for Pr$_3$ScO$_6$ and 0.92 for Pr$_3$Sc$_5$O$_{12}$ compounds. At a photon energy around 7 eV, the refractive index $n(\omega)$ achieves its maximum value. After that, the energy level gradually decreases until it reaches its lowest point, after which it hardly changes at all in the high energy zone ($\geq$50 eV).
\section{Conclusion}
In conclusion, this work presents a novel approach to discover new Pr-based perovskite oxide materials with desirable electronic and optical properties for photoluminescence applications. The use of a ML approach to screen a large number of candidate materials followed by DFT calculations to confirm their potential allowed for a more efficient and comprehensive approach to materials discovery and design. The predicted compounds (Pr$_3$AlO$_6$, Pr$_4$Al$_2$O$_9$, Pr$_3$ScO$_6$ and Pr$_3$Sc$_5$O$_{12}$) in the Pr-Al-O and Pr-Sc-O compound families were evaluated for their band structure, mechanical stability, density of states, optical absorption and emission spectra, confirming their potential for photoluminescence applications. \par Compared to their halide counterparts, perovskite oxide derivatives are often more thermally stable and less toxic, which makes them more suitable for practical applications. The perovskite oxide derivatives have  diverse design space, which allows for the incorporation of a wide range of luminescent and  enables the tuning of their electronic and optical properties. In addition, the chemical space of perovskite oxide derivatives is not fully explored yet, offering a promising avenue for the development of new luminescent materials. So, this work provides insights for future experimental investigations and can lead to the development of new materials for a variety of technological applications.
\section*{Acknowledgement}
\noindent This research was supported by National R\&D Program through the National Research  Foundation of Korea (NRF) funded by Ministry of Science and ICT (and RS-2023-00209910). Upendra Kumar expresses sincere gratitude to Dr. Sanjay Nayak, a Postdoctoral Fellow at Linköping University, for providing valuable motivation and inspiration to pursue work in the field of machine learning.
\enlargethispage{9 \baselineskip}
\section*{Author contributions}
\noindent Upendra Kumar and Hyeon Woo Kim conceived the idea and contributed equally to this project. Sobhit Singh provided key suggestions for manuscript modifications. Upendra Kumar wrote the manuscript and all authors read and reviewed it. Hyunseok Ko and Sung Beom Cho supervised the project. 
\newpage
\newcommand{\beginsupplement}{%
	\setcounter{table}{0}
	\renewcommand{\thetable}{S\arabic{table}}%
	\setcounter{figure}{0}
	\renewcommand{\thefigure}{S.F-\Roman{figure}}%
	\renewcommand\thesection{S.\arabic{section}}
	\def\thesection{\alph{section}}
	\renewcommand\thesubsection{\thesection.\Alph{subsection}}
	\renewcommand{\theequation}{S.\arabic{equation}}
}

\section*{Supplementary Material}\label{Upendra}
\beginsupplement
\section*{I. Data Mining}\label{}
\subsection*{I:A. Cr Based Ternary Perovskites Oxide}
\begin{table}[h]
\scalebox{0.55}{\begin{tabular}
{|c|c|c|c|c|c}
\hline
&&&\\
{} & \quad Pretty Formula \quad & \quad Bandgap (eV) \quad & \quad Debye Temp. (K)  \quad \\
&&&\\
\hline
&&& \\
1  &      Al$_{19}$CrO$_{30}$ &    3.92 &         912.37  \\
2   &        Al$_3$CrO$_6$ &    3.82 &         826.35  \\
3  &        Al$_3$CrO$_6$ &    3.59 &         833.75  \\
4   &       Cr(PO$_3$)$_3$ &    3.56 &         529.27 \\
5   &        Cr(CO)$_6$ &    3.46 &         197.17 \\
6  &       Cr(PO$_3$)$_3$ &    3.33 &         504.34\\
7  &       Cr(PO$_3$)$_3$ &    3.29 &         537.50  \\
8  &  Cs$_2$CrO$_4$ &    3.25 &         147.79  \\
9   &       Cr$_2$P$_4$O$_{13}$ &    3.25 &         558.18  \\
10   &       Cr(PO$_3$)$_3$ &    3.22 &         529.67 \\
11  &       Cr(PO$_3$)$_3$ &    3.21 &         533.14  \\
12  &         LiCrO$_2$ &    3.19 &         706.65  \\
13  &       Cr(PO$_3$)$_3$ &    3.18 &         525.11  \\
14  &        Na2CrO$_4$ &    3.15 &         282.81  \\
15  &        Rb2CrO$_4$ &    3.14 &         206.22  \\
16  &         K$_2$CrO$_4$ &    3.12  &         245.40   \\
17 &       Cr(PO$_3$)$_3$ &    3.10 &         564.44\\
18  &       Cr(PO$_3$)$_3$ &    3.10 &         561.24  \\
19  &         K$_2$CrO$_4$ &    3.08 &         260.97  \\
20 &        Li$_2$CrO$_4$ &    3.07 &         385.98  \\
\hline
\end{tabular}}
\caption{\small The table of compounds has the highest value of Debye temperature in the Cr-based ternary perovskites oxide family.}
\end{table}
\subsection*{I:B. Eu Based Ternary Perovskites Oxide}
\begin{table}[h]
\scalebox{0.55}{
\begin{tabular}
{|c|c|c|c|c|c}
\hline
&&&\\
{}  &\quad Pretty Formula \quad & \quad Bandgap (eV) \quad & \quad Debye Temp. (K)  \quad \\
&&&\\
\hline
&&& \\
1   &         EuB$_4$O$_7$ &    2.18 &         677.98  \\
2   &         Eu$_4$I$_6$O &    1.21 &         123.20  \\
3   &          EuSO$_4$ &    1.09 &         390.47 \\
4   &        Eu$_4$Cl$_6$O &    0.99 &         207.37  \\
5   &        Eu$_4$Br$_6$O &    0.85 &         177.82  \\
6   &        EuP$_5$O$_{14}$ &    0.77 &         493.42  \\
7   &         EuHfO$_3$ &    0.60 &         425.88 \\
8  &      Eu(ReO$_4$)$_2$ &    0.49 &         124.10  \\
9   &         EuY$_2$O$_4$ &    0.49 &         457.41  \\
10   &         EuY$_2$O$_4$ &    0.43 &         427.90  \\
11  &       Eu(PO$_3$)$_3$ &    0.42 &         426.65  \\
12  &        EuHo$_2$O$_4$ &    0.41 &         376.59  \\
13  &         EuZrO$_3$ &    0.40 &         463.58  \\
14  &         EuZrO$_3$ &    0.39 &         481.29  \\
15  &      Eu$_3$(BO$_3$)$_2$ &    0.38 &         388.91  \\
16 &         EuZrO$_3$ &    0.34 &         456.98  \\
17  &          EuCO$_3$ &    0.33 &         368.93\\
18  &       Eu(BO$_2$)$_2$ &    0.29 &         409.93  \\
19  &        Eu$_2$SiO$_4$ &    0.29 &         386.53  \\
20 &       Eu$_2$Al$_3$O$_8$ &    0.26 &         397.95  \\
&&&\\
\hline
\end{tabular}}
\caption{\small The table of compounds has the highest value of Debye temperature in the Eu-based ternary perovskites oxide family.}
\end{table}
\newpage 
\subsection*{I:C. Yb Based Ternary Perovskites Oxide}
\begin{table}[h]
\scalebox{0.55}{
\begin{tabular}
{|c|c|c|c|c|c}
\hline
&&&\\
{}  &\quad Pretty Formula \quad & \quad Bandgap (eV) \quad & \quad Debye Temp. (K)  \quad \\
&&&\\
\hline
&&& \\

1  &         YbB$_4$O$_7$ &    7.00 &         708.64  \\
2   &        Yb$_4$Cl$_6$O &    5.20 &         228.37  \\
3   &          YbCO$_3$ &    4.72 &         385.30  \\
4   &         YbHfO$_3$ &    4.52 &         397.59  \\
5   &         YbSiO$_3$ &    4.47 &         634.14  \\
6   &        Yb$_4$Br$_6$O &    4.25 &         205.44  \\
7   &          YbCO$_3$ &    4.13 &         415.56  \\
8   &         YbTiO$_3$ &    3.58 &         447.01  \\
9   &         YbMoO$_4$ &    3.36 &         350.55  \\
10   &        Yb$_4$Br$_6$O &    3.01 &         204.48  \\
11  &        K$_2$Yb$_2$O$_3$ &    2.56 &         270.51  \\
12  &         YbCrO$_4$ &    2.33 &         366.56  \\
13  &         YbTiO$_3$ &    2.30 &         514.07  \\
14  &       Yb$_2$Al$_2$O$_5$ &    2.28 &         515.06  \\
15  &         YbTiO$_3$ &    1.96 &         531.43 \\
16 &      Yb$_2$Te$_5$O$_{13}$ &    1.83 &         289.02  \\
17 &         YbCeO$_3$ &    1.81 &         368.72  \\
18  &          YbI$_2$O &    1.79 &         139.81  \\
19 &         YbSnO$_3$ &    1.77 &         442.12  \\
20  &        Yb$_2$Br$_2$O &    1.53 &         300.40  \\
&&&
\\
\hline
\end{tabular}}
\caption{\small The table of compounds has the highest value of Debye temperature in the Yb-based ternary perovskites oxide family.}
\end{table}
\FloatBarrier
\subsection*{I:D. Pr Based Ternary Perovskites Oxide}
\begin{table}[h]
\scalebox{0.55}{
\begin{tabular}
{|c|c|c|c|c|c}
\hline
&&&\\
{}  &\quad Pretty Formula \quad & \quad Bandgap (eV) \quad & \quad Debye Temp. (K)  \quad \\
&&&\\
\hline
&&& \\
1   &          PrPO$_4$ &      5.65 &           458.60  \\
2   &       Pr(BO$_2$)$_3$ &      5.50 &           498.45  \\
3   &          PrPO$_4$ &      5.49 &           415.15  \\
4   &       Pr(PO$_3$)$_3$ &      5.49 &           430.51  \\
5   &        PrP$_5$O$_{14}$ &      5.39 &           456.38  \\
6   &      Pr$_2$(SO$_4$)$_3$ &      5.39 &           337.50  \\
7   &       Pr$_2$Si$_2$O$_7$ &      5.17 &           413.46  \\
8   &      Pr$_4$B$_{10}$O$_{21}$ &      4.93 &           566.54 \\
9   &      Pr(ClO$_4$)$_3$ &      4.90 &           289.73  \\
10   &         PrLuO$_3$ &      4.88 &           379.47  \\
11  &       Pr$_2$Si$_2$O$_7$ &      4.84 &           442.59  \\
12  &          PrClO &      4.70 &           307.96  \\
13 &          PrBO$_3$ &      4.67 &           437.17 \\
14 &          PrBrO &      4.46 &           256.43  \\
15  &         PrScO$_3$ &      4.26 &           513.98  \\
16  &         PrAlO$_3$ &      4.22 &           609.67  \\
17 &        Pr(HO)$_3$ &      4.16 &           421.41  \\
18  &        Pr(HO)$_3$ &      4.11 &           419.38 \\
19  &      Pr2(WO$_4$)$_3$ &      3.98 &           301.21  \\
20  &         PrNbO$_4$ &      3.86 &           402.33  \\
&&&\\
\hline
\end{tabular}}
\end{table}
Sc and Al are very under explored families in the Pr-chalcogen-O compounds, as mentioned in Material Project\footnote{A. Jain et al., ``Commentary: The materials project: A materials genome approach to accelerating materials innovation,” APL Mater. \textbf{1}, 011002(2013).\\
\url{https://doi.org/10.1063/1.4812323}}. The Pr-based ternary perovskites oxides are mentioned in Table(\ref{PrO_Family_Tab}).
\begin{table}[h]
\centering
\scalebox{0.85}{\begin{tabular}{|c|c|c|c|c|c|c|}
\hline
{} & Material Project ID & Pretty Formula &  Band gap &  Debye Temp. & Ehull (eV) \\
\hline
1 &   mp-559756 &         PrScO$_3$ &    4.26 &         513.98 & \quad    0 \\
2 &     mp-3466 &         PrAlO$_3$ &    4.22 &         609.67 & \quad    0 \\
\hline
\end{tabular}}
\caption{\small The value of bandgap and Debye temperature of Pr-chalcogen-O compounds.}
\label{PrO_Family_Tab}
\end{table}
\section*{II. Convex Hull}
\begin{table}[h]
\centering
\scalebox{0.95}{\begin{tabular}{|c|c|c|c|c|c|c|}
\hline
\thead{\textbf{Parent Atom} \\ {}} &\thead{\textbf{Space Group}\\ \textbf{Type}}& \thead{\textbf{Crystal}\\ \textbf{Structures}} & \thead{\textbf{Space Group}\\ {\textbf{Number}}}\\
\hline
&&&\\
Pr & \quad  $P63/mmc$  \quad &  \quad Hexagonal \quad& 194\\
&&&\\
\hline
&&&\\
Al& $Fm3m$  & Cubic & 225\\
&&&\\
\hline
&&&\\
Sc& $P63/mmc$  & Hexagonal & 194\\
&&&\\
\hline
&&&\\
O & $C2/m$  & Monoclinic & 12\\
&&&\\
\hline
\end{tabular}}
\caption{\small The details of parent atom space group and crystal structure.}
\label{parent_atom}
\end{table}
\begin{table}[h]
\centering
\scalebox{0.85}{
\begin{tabular}{|c|c|c|c|c|c|c|}
\hline
&&&&&& \\
& \textbf{Composition} &        \textbf{Space Group} & \textbf{Crystal System} &       \textbf{Ehull (eV/atom)} & \thead{\textbf{ Band Gap(eV)} } &  \textbf{Direct} \\
& &&&&& \\
\hline
& &&&&& \\
1& Pr$_4$Al$_2$O$_9$ &     (P$2_1$/c, 14) &     monoclinic &    0 &    4.07 &    True \\
2& PrAlO$_3$ &       (Pnma, 62) &   orthorhombic &    0&    3.72 &   False \\
3& Pr$_3$AlO$_6$ &     (Cmc21, 36) &   orthorhombic &    0 &    4.17 &   False \\
4& Pr$_{4}$Al$_{2}$O$_{9}$ &     (P2$_1$/c, 14) &     monoclinic &    1.96 &    4.18 &   False \\
5& Pr$_{3}$Al$_{5}$O$_{12}$ &     ('Ia$\bar{3}$d', 230) &          cubic &   10.80 &    3.76 &    True \\
6& PrAlO$_{3}$ &       (Cmcm, 63) &   orthorhombic &   16.74 &    4.25 &    True \\
7& PrAlO$_{3}$ &         (P$\bar{1}$, 2) &      triclinic &   42.40 &    4.29 &   False \\
8& PrAlO$_{3}$ &     (P$2_1$/c, 14) &     monoclinic &   43.39 &    4.25 &   False \\
9& PrAl3O$_{6}$ &       (C2/c, 15) &     monoclinic &   48.50 &    4.19 &   False \\
10& PrAlO$_{3}$ &         ('P$\bar{1}$', 2) &      triclinic &   48.91 &    4.23 &   False \\
11& PrAlO$_{3}$ &     (P2$_1$/c, 14) &     monoclinic &   61.80 &    4.45 &   False \\
12& PrAl$_{11}$O$_{18}$ &         (P$\bar{1}$, 2) &      triclinic &   67.48 &    2.63 &   False \\
13& Pr$_{4}$Al$_{6}$O$_{15}$ &       (C2/c, 15) &     monoclinic &   79.38 &    4.21 &   False \\
14& Pr$_{2}$Al$_{4}$O$_{9}$ &       (Pbam, 55) &   orthorhombic &   81.72 &    3.35 &   False \\
15& PrAlO$_{3}$ &     (P2$_1$/c, 14) &     monoclinic &   87.84 &    4.33 &    True \\
16& PrAlO$_{3}$ &         (P$\bar{1}$, 2) &      triclinic &   94.27 &    4.26 &    True \\
17& Pr$_{3}$AlO$_{6}$ &       (R$\bar{3}$, 148) &       trigonal &  110.70 &    4.01 &   False \\
18& PrAlO$_{3}$ &       (Pnma, 62) &   orthorhombic &  122.13 &    3.51 &    True \\
19& PrAlO$_{3}$ &         (P$\bar{1}$, 2) &      triclinic &  128.87 &    3.02 &   False \\
20& PrAlO$_{3}$ &  (P6$_3$/mmc, 194) &      hexagonal &  209.10 &    3.22 &   False \\
& &&&&& \\
\hline
\end{tabular}
}
\caption{\small Predicted structures in the  Pr-Al-O ternary compound family.}
\label{Pr-Al-O-Predicted-Str}
\end{table}
\begin{table}[h]
\centering
\scalebox{0.85}{\begin{tabular}{|c|c|c|c|c|c|c|}
\hline
&&&&&& \\
&{}              \textbf{Composition} &        \textbf{Space Group} & \textbf{Crystal System} &       \textbf{Ehull (eV/atom)} & \thead{\textbf{ Band Gap(eV)} } &  \textbf{Direct} \\
&&&&&& \\
\hline
&&&&&& \\
1 &      PrScO$_{3}$ &      (R$\bar{3}$c, 167) &       trigonal &    0 &    3.87 &    True \\
2  &   Pr$_{3}$Sc$_{5}$O$_{12}$ &     (Ia$\bar{3}$d, 230) &          cubic &    0 &    3.99 &    True \\
3   &     Pr$_{3}$ScO$_{6}$ &       (R$\bar{3}$, 148) &       trigonal &   0 &    4.38 &   False \\
4  &     PrSc$_{3}$O$_{6}$ &       (C12/c, 15) &     monoclinic &   26.35 &    3.99 &   False \\
5  &     Pr$_{3}$ScO$_{6}$ &     (Cmc2$_1$, 36) &   orthorhombic &   33.49 &    3.94 &   False \\
6  &      PrScO$_{3}$&     (Pmc2$_1$, 26) &   orthorhombic &   33.72 &    3.49 &    True \\
7   &      PrScO$_{3}$ &  (P6$_3$/mmc, 194) &      hexagonal &   33.83 &    3.27 &    True \\
8 &    PrSc$_{7}$O$_{12}$ &      (Im$\bar{3}$, 204) &          cubic &   43.66 &    3.85 &    True \\
9  &    Pr$_{4}$Sc$_{2}$O$_{9}$ &     (P2$_1$/c, 14) &     monoclinic &   63.88 &    3.46 &   False \\
10  &    Pr$_{4}$Sc$_{2}$O$_{9}$ &     (P2$_1$/c, 14) &     monoclinic &   65.57 &    3.41 &   False \\
11  &      PrScO$_{3}$ &    (P4/mbm, 127) &     tetragonal &   73.44 &    2.63 &    True \\
12  &      PrScO$_{3}$ &         (P$\bar{1}$, 2) &      triclinic &   86.83 &    3.69 &   False \\
13  &      PrScO$_{3}$ &     (P2$_1$/c, 14) &     monoclinic &   92.35 &    3.81 &   False \\
14  &    Pr$_{2}$Sc$_{4}$O$_{9}$ &       (Pbam, 55) &   orthorhombic &   92.66 &    3.54 &    True \\
15  &      PrScO$_{3}$ &         (P$\bar{1}$, 2) &      triclinic &  102.30 &    3.51 &   False \\
16  &      PrScO$_{3}$ &     (P2$_1$/c, 14) &     monoclinic &  105.05 &    3.54 &   False \\
17  &      PrScO$_{3}$ &         (P$\bar{1}$, 2) &      triclinic &  106.98 &    3.69 &   False \\
18  &      PrScO$_{3}$ &          (P1, 1) &      triclinic &  115.35 &    3.77 &   False \\
19  &      PrScO$_{3}$ &          (P1, 1) &      triclinic &  152.51 &    2.62 &    True \\
20  &   Pr$_{4}$Sc$_{6}$O$_{15}$ &       (C2/c, 15) &     monoclinic &  176.94 &    3.62 &   False \\
21  &      PrScO$_{3}$ &       (Pnma, 62) &   orthorhombic &  191.75 &    3.66 &   False \\
22  &      PrScO$_{3}$ &     (Pm$\bar{3}$m, 221) &          cubic &  206.52 &    2.41 &    True  \\
23  &      PrScO$_{3}$ &       (C2/m, 12) &     monoclinic &  211.37 &    3.22 &   False \\
24  &      PrScO$_{3}$ &     (P2$_1$/c, 14) &     monoclinic &  299.50 &    2.43 &   False \\
&&&&&& \\
\hline
\end{tabular}}
\caption{\small Predicted structures in the  Pr-Sc-O ternary compound family.}
\label{Pr-Sc-O-Predicted-Str}
\end{table}
\FloatBarrier
\section*{III. Mechanical Properties} 
To ensure that a material is properly included into the developing technology, it is essential to extensively analyse its mechanical characteristics. By applying the Born criterion, we assess the mechanical stability\footnote{F. Mouhat and F.-X. Coudert, ``Necessary and sufficient elastic stability conditions in various crystal systems,” Phys. Rev. B \textbf{90}, 224104(2014). \\ \url{https://doi.org/10.1103/PhysRevB.90.224104}}. The elastic tensor matrix is obtained using finite lattice distortions and the strain-stress relationship. In order to account for the relaxation of rigid ions, the elastic tensor matrix $(\mathrm{C}_{ij})$ has been computed. The coefficients are measured in gigapascals (GPa). The \texttt{MechElastic Python module}\footnote{S. Singh et al.,
``Mechelastic: A python library for analysis of mechanical and elastic properties of bulk and 2d
materials,” Comput. Phys. Commun.,  \textbf{267}, 108068(2021). \\
\url{https://doi.org/10.1016/j.cpc.2021.108068}} is used for the mechanical characteristics of newly predicted compounds by utilising the elastic coefficient matrix ($\mathrm{C}_{ij}$) generated from any ab-initio density-functional theory (DFT) method. 
\begin{table}[h]
\centering
\scalebox{0.85}{\begin{tabular}{|c|c|c|c|c|}
\hline
\quad \thead{ \\ \textbf{ Parameters} \\ {}} \quad& \quad \thead{ \\ \textbf{ Pr$_3$AlO$_6$ } \\ {} }  \quad &  \quad \thead{\\ \textbf{Pr$_4$Al$_2$O$_9$} \\ {}}   \quad &  \quad \thead{\\ \textbf{Pr$_3$ScO$_6$} \\ {} }  \quad &  \quad \thead{ \\ \textbf{Pr$_3$Sc$_5$O$_{12}$} \\ {}}  \quad \\
\hline
\thead{ \\ Bulk  \\ {modulus   (GPa)} }  & \thead{ \\121.19 \\ {}} &  \thead{ \\ 117.97 \\ {}} & \thead{ \\128.86 \\{}} & \thead{ \\128.84 \\ {}} \\
\hline
\thead{ \\ Shear  \\ {modulus  (GPa)} } & \thead{ \\52.37 \\ {}} &  \thead{ \\56.60\\ {}} & \thead{ \\50.45\\{}} &  \thead{ \\ 61.48 \\ {}}\\
\hline
\thead{ \\  Young's  \\ {modulus  (GPa)} } & \thead{ \\ 137.33 \\{} }  & \thead{ \\146.39\\{}} & \thead{ \\133.86\\{}} & \thead{ \\ 159.14\\ {}} \\
\hline
\thead{ \\ Poisson's  \\ {ratio} } & \thead{ \\ 0.31 \\{}} & \thead{ \\0.29 \\ {}} & \thead{ \\0.33\\{}} & \thead{ \\0.29 \\ {}} \\
\hline
\end{tabular}}
\caption{\small Calculated material properties of predicted structures.}
\label{mech_properties}
\end{table}
The elastic tensor matrices for Pr$_3$AlO$_6$ has form:
\begin{align}
\mathrm{C}_{ij} (\mathrm{GPa})=
\begin{bmatrix}
204.85 &   67.26   & 84.16 &     0  &  0 &      0 \\
67.26 & 185.02 &   85.07 &     0 &    0 &      0 \\
84.16 &   85.07 &  237.75  &    0    & 0   &    0 \\ 
0  &   0  &    0 &    52.50  &   0  &   0 \\
0 &    0  &   0  &   0  &  45.87  &    0 \\
0  &    0   &  0  &   0 &    0 &   38.66
\end{bmatrix}
\end{align}
In the case of Pr$_3$AlO$_6$, which is part of the orthorhombic crystal system, the essential conditions are as follows:
\begin{enumerate}
\item $\mathrm{C}_{11}  >  0$
\item $\mathrm{C}_{11}\times \mathrm{C}_{22} \ > \mathrm{C}_{12}^2$
\item $\mathrm{C}_{11}\times \mathrm{C}_{22}\times \mathrm{C}_{33} \ + \ 2\mathrm{C}_{12}\times \mathrm{C}_{13}\times \mathrm{C}_{23} \ - \ \mathrm{C}_{11}\times \mathrm{C}_{23}^2 \ - \ \mathrm{C}_{22}\times \mathrm{C}_{13}^2 \ -  \mathrm{C}_{33}\times \mathrm{C}_{12}^2 > 0$
\item $\mathrm{C}_{44}  > 0$
\item $\mathrm{C}_{55}  > 0 $
\item $\mathrm{C}_{66}  > 0 $	
\end{enumerate}
The elastic tensor matrices for Pr$_4$Al$_2$O$_9$ has form:
\begin{align}
\mathrm{C}_{ij} (\mathrm{GPa})=
\begin{bmatrix}
205.77  &   83.24   &  82.13   &   2.68   &  0  &   0 \\
83.24  &  195.80  &   81.60  &   -2.54  &    0  &  0 \\
82.13 &     81.60 &    169.47  &   -1.73 &     0 &     0 \\
2.68  &   -2.54 &    -1.73 &    54.80 &    0 &     0
\\
0  &    0 &    0  &   0 &   61.25   &  -1.83 \\
0  &   0 &    0 &    0  &   -1.83  &   61.43  
\end{bmatrix}
\end{align}
For Pr$_4$Al$_2$O$_9$ which belongs to the monoclinic crystal system, the necessary criteria are given as;
\begin{enumerate}
\item{C$_{11}$ > 0} 
\item{C$_{22}$ > 0} 
\item{C$_{33}$ > 0} 
\item{C$_{44}$ > 0} 
\item{C$_{55}$ > } 
\item{C$_{66}$ > 0} 
\item{$\big[$C$_{11}$ + C$_{22}$ + C$_{33}$ + 2$\times$(C$_{12}$ + C$_{13}$ + C$_{23}$)$\big]$ > 0} 
\item{C$_{33}$$\times$C$_{55}$ - C$_{35}^2$ > 0} 
\item{C$_{44}$$\times$C$_{66}$ - C$_{46}^2$ > 0 (x) C$_{22}$ + C$_{33}$ - 2$\times$C$_{23}$  > 0 } 
\item{C$_{22}$$\times$(C$_{33}$$\times$C$_{55}$ - C$_{35}$$^2$) + 2$\times$C$_{23}$$\times$C$_{25}$$\times$C$_{35}$ - (C$_{23}$)$^2$$\times$C$_{55}$ - (C$_{25}$)$^2$$\times$C$_{33}$   > 0}
\item{2$\times$[C$_{15}$$\times$C$_{25}$$\times$(C$_{33}$$\times$C$_{12}$ - C$_{13}$$\times$C$_{23}$) + C$_{15}$$\times$C$_{35}$$\times$(C$_{22}$$\times$C$_{13}$ - C$_{12}$$\times$C$_{23}$) + C$_{25}$$\times$C$_{35}$$\times$(C$_{11}$$\times$C$_{23}$ - C$_{12}$$\times$C$_{13}$)] - [C$_{15}$$\times$C$_{15}$$\times$(C$_{22}$$\times$C$_{33}$ - C$_{23}$$^2$) + C$_{25}$$\times$C$_{25}$$\times$(C$_{11}$$\times$C$_{33}$ - C$_{13}$$^2$) + C$_{35}$$\times$C$_{35}$$\times$(C$_{11}$$\times$C$_{22}$ - C$_{12}$$^2$)] + C$_{55}$$\times$g > 0        
where, g = [C$_{11}$$\times$C$_{22}$$\times$C$_{33}$ - C$_{11}$$\times$C$_{23}$$\times$C$_{23}$ - C$_{22}$$\times$C$_{13}$$\times$C$_{13}$ - C$_{33}$$\times$C$_{12}$$\times$C$_{12}$ + 2$\times$C$_{12}$$\times$C$_{13}$$\times$C$_{23}$ ].
} 
\end{enumerate}
The elastic tensor matrices for Pr$_3$ScO$_6$ has form:
\begin{align}
\mathrm{C}_{ij} (\mathrm{GPa})= \begin{bmatrix}
185.28&     92.99  &   98.50   &  -4.21  &  -14.20  &    6.47 \\
92.99   &  214.09  &   93.92   &  -4.96   &   7.98   &  -5.98 \\
98.50  &   93.92  &  195.43   &   8.53  &  -11.13  &   -0.95 \\
-4.21  &   -4.96   &   8.53  &   60.74   &  -0.29   &  -8.01  \\
-14.20   &   7.98 &  -11.13 &    -0.29   &  45.79   &  -4.37 \\
6.47 &    -5.98 &    -0.95  &   -8.01  &   -4.37 &    50.59
\end{bmatrix}
\end{align}
For Pr$_3$ScO$_6$ which belongs to the rhombohedral-2 crystal system, the necessary criteria are given as;
\begin{enumerate}
\item{C$_{11}$ - C$_{12}$ > 0} 
\item{C$_{13}$$^2$ < (1/2)$\times$C$_{33}$(C$_{11}$ + C$_{12}$)}
\item{C$_{14}$$^2$ + C$_{15}$$^2$ < (1/2)$\times$C$_{44}$$\times$(C$_{11}$-C$_{12}$) = C$_{44}$$\times$C$_{66}$}
\item{C$_{44}$ > 0}
\end{enumerate}
The elastic tensor matrices for Pr$_3$Sc$_5$O$_{12}$ has form:
\begin{align}
\mathrm{C}_{ij} (\mathrm{GPa})= \begin{bmatrix}
222.24 &     82.08   &   82.08   &  0  &   0    &  0 \\
82.081 &    222.24  &   82.081 &      0   &    0 &     0 \\
82.081 &    82.081 &    222.24 &      0 &      0  &     0 \\
0 &     0 &      0 &     56.54  &     0  &   0 \\
0  &    0 &    0  &    0 &     56.54  &     0  \\
0  &     0  &     0 &    0 &     0  &   56.54 
\end{bmatrix}
\end{align}
For Pr$_3$Sc$_5$O$_{12}$ which belongs to the cubic crystal system, the necessary criteria are given as;
\begin{enumerate}
\item{C$_{11}$ - C$_{12}$ > 0} 
\item{C$_{11}$ + 2C$_{12}$ > 0}
\item{C$_{44}$ > 0}
\end{enumerate}
All of the Born criterion are met by the coefficient produced using DFT-PBE, indicating that all predicted structures are mechanically stable. Other mechanical properties-related parameters are listed in Table(\ref{mech_properties}).

\end{document}